\documentclass[twocolumn,floatfix,showpacs
]{revtex4}

\usepackage{amsfonts}
\usepackage{amsmath}
\usepackage{amssymb}
\usepackage{mathrsfs}
\usepackage[english,activeacute]{babel}
\usepackage[latin1]{inputenc}
\usepackage{graphicx}
\usepackage{subfigure}
\usepackage{epsfig}
\usepackage{float}

 \usepackage{color}

\begin{document}

\title{Planar motion with Fresnel integrals as components of the velocity}

\author{Elizabeth Flores-Gardu\~no}\email{elizabeth.flores@ipicyt.edu.mx}
\affiliation{IPICyT, Instituto Potosino de Investigacion Cientifica y Tecnologica,\\
Camino a la presa San Jos\'e 2055, Col. Lomas 4a Secci\'on, 78216 San Luis Potos\'{\i}, S.L.P., Mexico}
\author{Stefan C. Mancas}\email{mancass.erau.edu} %$^{\diamond}$}
\affiliation{Department of Mathematics, Embry-Riddle Aeronautical University, Daytona Beach, FL 32114-3900, USA}
\author{Haret C. Rosu}\email{hcr@ipicyt.edu.mx}
\affiliation{IPICyT, Instituto Potosino de Investigacion Cientifica y Tecnologica,\\
Camino a la presa San Jos\'e 2055, Col. Lomas 4a Secci\'on, 78216 San Luis Potos\'{\i}, S.L.P., Mexico}
\author{Maximino P\'erez-Maldonado}\email{maximino.perez@upslp.edu.mx}
\affiliation{Universidad Polit\'ecnica de San Luis Potos\'{\i},\\
Urbano Villal\'on No 500 Col. La Ladrillera C.P 78363 San Luis Potos\'{\i}, S.L.P., Mexico}

%ode's; Newtonian mechanics

\begin{abstract}
\noindent We analyze the two-dimensional motion of a rigid body due to a constant torque generated by a force acting on the body parallel to the surface on which the body moves extending an old note of Ferris-Prabhu [Am. J. Phys. \textbf{38}, 1356-1357 (1970)] and supplementing it with a short discussion of the jerking properties. \\


%\bigskip
%
%\noindent  Analizamos el movimiento dos dimensional de un cuerpo rigido bajo la acci\'on de una torca constante generada por una fuerza aplicada al cuerpo paralelamente a la superficie sobre la cual se mueve. \\
%
%\noindent {\bf Descriptores}: movimiento plano; torca; integrales de Fresnel; aceleraci\'on dependiente del tiempo.
\end{abstract}

\maketitle

\section{Introduction}

While there are recent surprising applications of Fresnel integrals, such as to rat whiskers \cite{ratw} and orange peel \cite{op},
it is surely much less known that fifty years ago Ferris-Prabhu \cite{Ferris} discussed
a two-dimensional motion of a rigid body in classical Newtonian mechanics as an example in which Fresnel integrals occur beyond their usual context of near-field optical diffraction generated by slits and apertures \cite{Hecht,Fowles,Rossi,Goodman,Iizuka}. Since the paper of Ferris-Prabhu is a very short note, and has also some ambiguous points, we provide here a more detailed analysis of the kinematical quantities of this interesting motion supported by their plots adding also a discussion of its jerked properties.

\section{Motion with velocity whose cartesian components are Fresnel integrals}

\begin{figure} [h!] \centering %[htb]
{\includegraphics[width=1.07\linewidth]{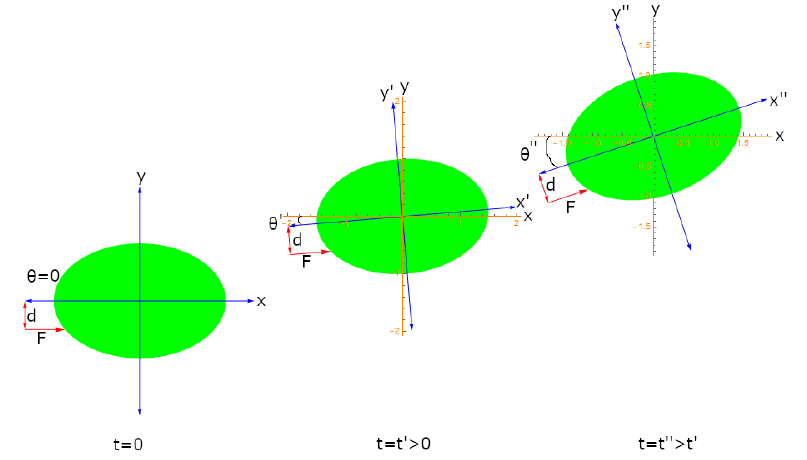}}
	\caption{ The planar motion of a rigid object considered in this paper at the initial moment and two subsequent instants.}
\label{f01}
\end{figure}

Let us consider a small rigid and compact object of mass $M$ %volume $V\sim D^3$,
and moment of inertia $I$ on a frictionless surface
defined by the cartesian coordinates $x$ and $y$ with the origin placed at the center of mass of the object.
We assume the object is initially at rest
and apply at time $t=0$ a constant force, $F$, along the positive $x$ direction at the point $(x,y)=(0,-d)$, where $d$ is some distance on the $y$ axis smaller than the size of the object in that direction. During the course of motion, the line of application of the force is maintained at the distance $d$ for any instantaneous angle $\theta$ made by the force with the $x$ axis, i.e., the force as a vector does not change in the rotating cartesian system defined by the axes $x'$ and $y'$ bound to the body, see Fig.~\ref{f01}.
We are interested in the trajectory of the center of mass of the object under these conditions. Choosing the center of mass as a reference point for the motion is theoretically very convenient because this planar motion is a superposition of translational and rotational motions and for the center of mass the translational motion is due to Newton's second law and the rotational motion is due to the torque equation in their standard form.

%.....FIGURE 01

\begin{figure} [h!] \centering %[htb]
{\includegraphics[width=0.49\linewidth]{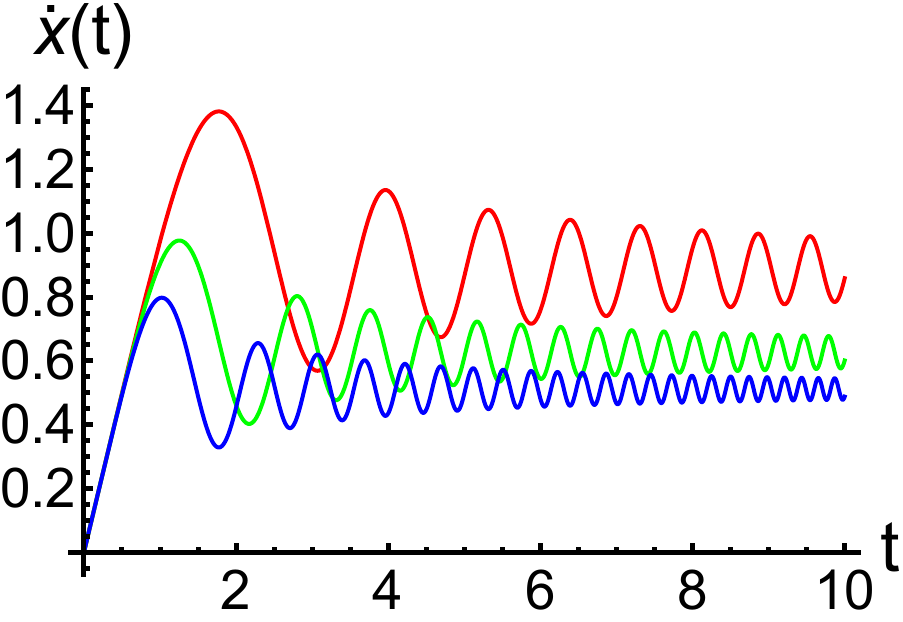}}
{\includegraphics[width=0.49\linewidth]{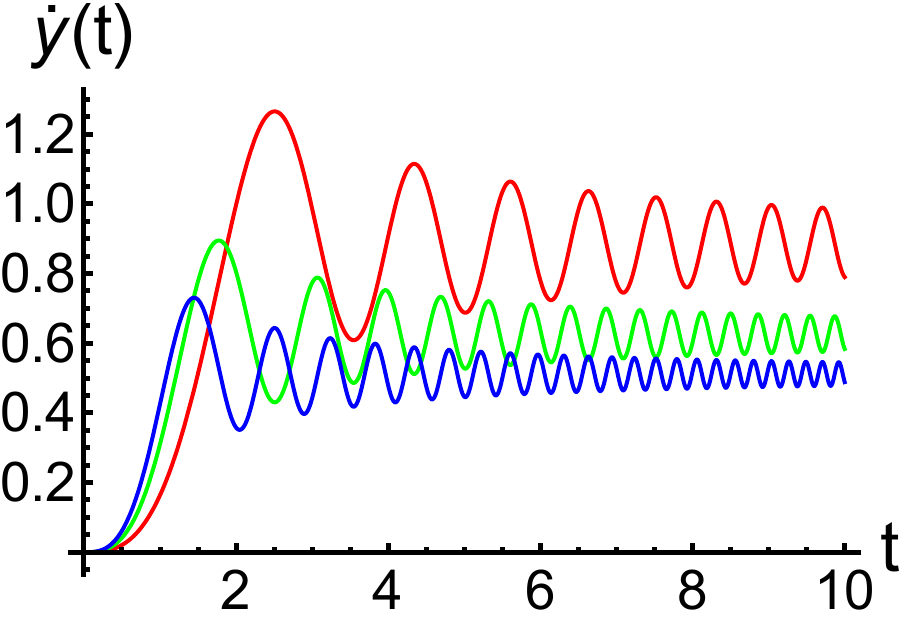}}
{\includegraphics[width=0.49\linewidth]{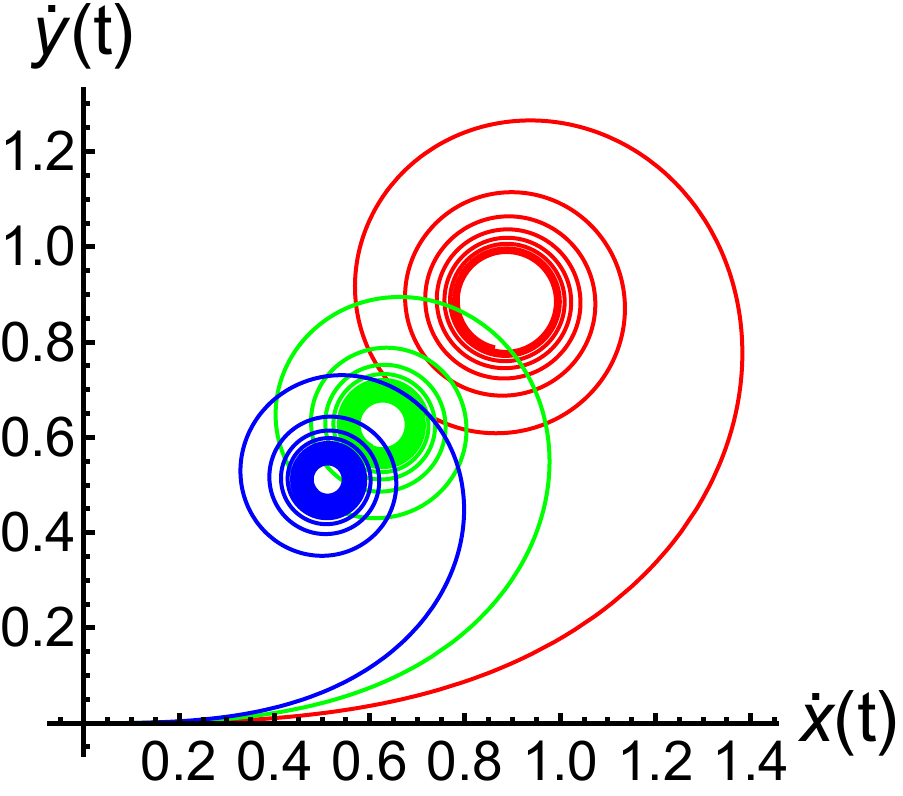}}
{\includegraphics[width=0.49\linewidth]{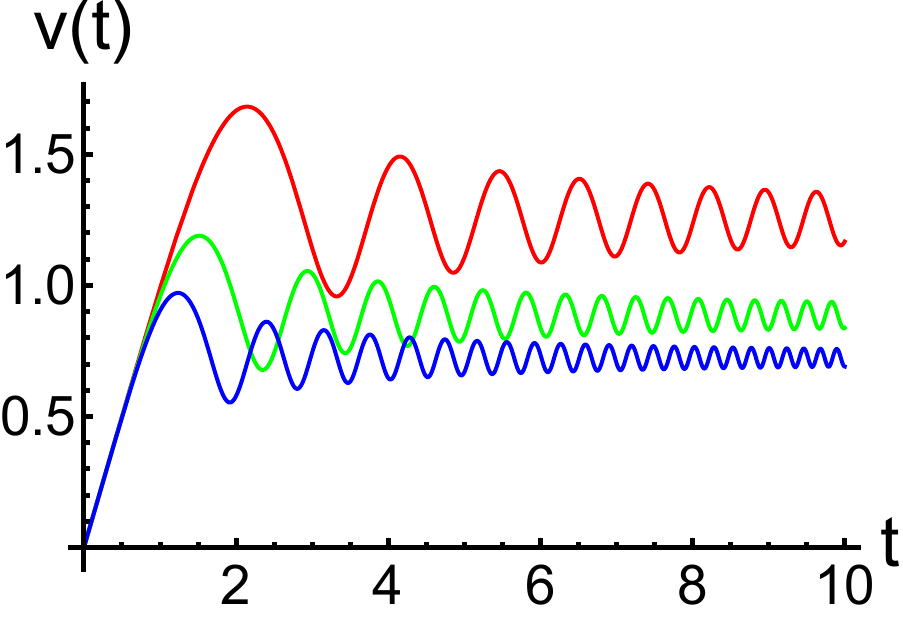}}
	\caption{ The cartesian components of the velocity for $K=1,2,3$ (red, green, and blue, respectively), and force and mass taken as unity.
The corresponding Argand plots (bottom left) and the speed $\mathrm{v}(t)=\sqrt{\dot{x}^2+\dot{y}^2}$ (bottom right).}
\label{vit}
\end{figure}

The torque equation for the motion as depicted in Fig.~\ref{f01} is
\begin{equation}\label{eq01}
\tau =I \ddot{\theta} = Fd~,
\end{equation}
where $I$ is the moment of inertia of the body and the dot stands for the time derivative.
Denoting $K =Fd/I$ and using zero initial conditions we find the polar angle
$\theta (t) = \frac{K}{2} t^{2}$.
We now use this quadratic angle to write the cartesian components of acceleration provided by Newton's second law
\begin{eqnarray}\label{eq1}
\begin{aligned}
	\ddot{x}(t) &=& \frac{F}{M} \cos \left( \frac{K}{2} t^{2} \right) ~, \\ %\label{eq3} \\
	\ddot{y}(t) &=& \frac{F}{M} \sin \left( \frac{K}{2} t^{2} \right) ~. %\label{eq4}
\end{aligned}
\end{eqnarray}
By one integration and assuming zero integration constants, we find the cartesian components of the velocity, expressed in terms of the Fresnel integrals, $C(\tau)=\int_0^\tau \cos(\pi u^2/2)du$ and $S(\tau)=\int_0^\tau \sin(\pi u^2/2)du$, where $\tau=\sqrt{\frac{K}{\pi}}\,t$ \cite{Ferris}
%.....eqs. 5,6
\begin{eqnarray}\label{eq2}
\begin{aligned}
\dot{x}(t) &=&  \frac{F}{M} \sqrt{\frac{\pi}{K}}\, C \left( \sqrt{\frac{K}{\pi }} t \right)  ~, \\ %\label{eq5} \\
\dot{y}(t) &=&  \frac{F}{M} \sqrt{\frac{\pi}{K}}\, S \left( \sqrt{\frac{K}{\pi }} t \right) ~. %\label{eq6}
\end{aligned}
\end{eqnarray}
Since both velocities are zero at $t=0$, it confirms that the center of mass undergoes a planar movement as an instant center of rotation.
These components are plotted in Fig.~\ref{vit} for three values of $K$, together with their renowned Argand plot
(the positive part of the clothoid/Cornu/Euler spiral \cite{rme}) and the speed $\mathrm{v}(t)=\sqrt{\dot{x}^2+\dot{y}^2}$.
%.....FIGURE 1

\medskip

Integrating again with zero integration constants, the cartesian components for the position are
%....eqs. 7,8
 {\small
 \begin{eqnarray}\label{eq3}
 \begin{aligned}
&x(t)=  \frac{F}{M} \sqrt{\frac{\pi}{K}} \left[ t\, C \left( \sqrt{\frac{K}{\pi}}t \right) - \frac{1}{\sqrt{\pi K}} \sin \left( \frac{K}{2} t^{2} \right) \right]  ~,\\ % \label{eq7} \\
&y(t) =  \frac{F}{M} \sqrt{\frac{\pi}{K}} \left[ t\, S \left( \sqrt{\frac{K}{\pi}}t \right) + \frac{1}{\sqrt{\pi K}} \left(\cos \left( \frac{K}{2} t^{2}\right)-1 \right) \right]. %\label{eq8}
\end{aligned}
\end{eqnarray}
}

Since $F/M$ is an overall scaling factor, we proceed here by assuming $F/M=1$, which does not change the analytical behavior of the solutions.
We plot the displacements (\ref{eq3}) in Fig.~\ref{fig7-8} together with the radial displacement $\mathrm{r}(t)=\sqrt{x^2+y^2}$.
The plots show a linear behavior which sets in at already moderate instants of time with some
superposed ripples which are smaller and almost disappearing at increasing $K$.
This is easy to understand by using the large argument expansion of the Fresnel integrals which we write in the form
{\small
\begin{eqnarray}\label{eq4}
\begin{aligned}
&C(t)\sim \frac{1}{2}{\rm sgn}(t)+\frac{1}{\pi t}\sin\left(\frac{\pi}{2}t^2\right)
%\qquad\quad\,\,
\sim \frac{1}{2}\left(1+t\,{\rm sinc}\left(\frac{\pi}{2}t^2\right)\right), \\
&S(t)\sim \frac{1}{2} {\rm sgn}(t)-\frac{1}{\pi t}\cos\left(\frac{\pi}{2}t^2\right)
%\qquad\quad \,\,
\sim \frac{1}{2}\left(1+t\,{\rm sinc}^{*}\left(\frac{\pi}{2}t^2\right)\right), %\label{asS}
\end{aligned}
\end{eqnarray}}
for $t\gg 1$ and where the notation
$$
{\rm sinc}^{*}\left(\frac{\pi}{2}t^2\right)=-\frac{\cos(\frac{\pi}{2}t^2)}{\frac{\pi}{2}t^2}=
\frac{\sin(\frac{\pi}{2}t^2-\pi/2)}{\frac{\pi}{2}t^2}~
$$
is introduced to emphasize the well-known sinc type ripples in the plateau region of the Fresnel integrals considered as switching functions. The linear rising in the amplitude of the ripples in the large asymptotic Fresnel integrals (\ref{eq4}) is by far compensated by the natural damping of the sinc oscillations.
In the case of the cartesian displacements (\ref{eq3}), we notice that the last oscillatory terms are bounded by their amplitude, $1/K$, and so their effect in the plots cannot be perceived.
%%% trayectoria del
In other words, the displacement plots are dominated by the even functions $t\,C$ and $t\,S$, which asymptotically in the first quadrant are given by
\begin{eqnarray}\label{eq5}
\begin{aligned}
&t\,C(t)&\sim \frac{1}{2}t\left(1 %\,\rm sgn}(t)
+t\,{\rm sinc}\left(\frac{\pi}{2}t^2\right)\right)~, \\ %\label{astC} \\
&t\,S(t)&\sim \frac{1}{2}t\left(1 %\,{\rm sgn}(t)
+t\,{\rm sinc}^{*}\left(\frac{\pi}{2}t^2\right)\right)~, %\label{astS}
\end{aligned}
\end{eqnarray}
to which the diminishing effect of the factor $\sqrt{\pi/K}$ should be added. Using (\ref{eq5}) in (\ref{eq3}) for $F/M=1$, one can see that the graphs of the displacements are essentially the straight lines $\sqrt{\pi/K}\,t/2$ with the superposed $1+t\, {\rm sinc}$ and $1+t\, {\rm sinc}^{*}$ modulations quickly damping down.

\medskip

\section{Some jerk properties}

We now point out some jerked properties of this kind of motion.
The cartesian displacements $x$ and $y$  present two coupled jerks of the type
\begin{eqnarray}\label{eq6}
\begin{aligned}
&\dddot{x}=-\widetilde{K}t\ddot{y}~, \\ %\label{jerk01}\\
& \dddot{y}=\widetilde{K}t\ddot{x}~, %\label{jerk02}
\end{aligned}
\end{eqnarray}
where $\widetilde{K}=FK/M$, as can be inferred from the derivatives of (\ref{eq1}), although the total acceleration is constant, $a=F/M$.
This is similar to the case of circular motion of arbitrary radius $R$ and angular velocity $\omega$, where the cartesian jerks are given by
 \begin{eqnarray}\label{jerk1}
 \begin{aligned}
&\dddot{x}=-\omega\ddot{y}~, \\ %\label{jerkc1}\\
&\dddot{y}=\omega\ddot{x}~, %\label{jerkc2}
\end{aligned}
\end{eqnarray}
and the centripetal acceleration is $a_c=\omega^2R$.

\medskip

It is interesting to find out the differential equation satisfied by the jerks, which {\em a priory} should be a third-order one \cite{B1910}.
Writing the system (\ref{eq3}) in the form
 \begin{eqnarray}\label{eq3new}
 \begin{aligned}
&x(t)= t \dot{x}-\frac{1}{K}\ddot{y}~,\\
&y(t)= t \dot{y}+\frac{1}{K}\ddot{x}-\frac{1}{K}~,
\end{aligned}
\end{eqnarray}
%%.............
%.....FIGURE 2
\begin{figure} [h!] \centering %[htb]
{\includegraphics[width=0.49\linewidth]{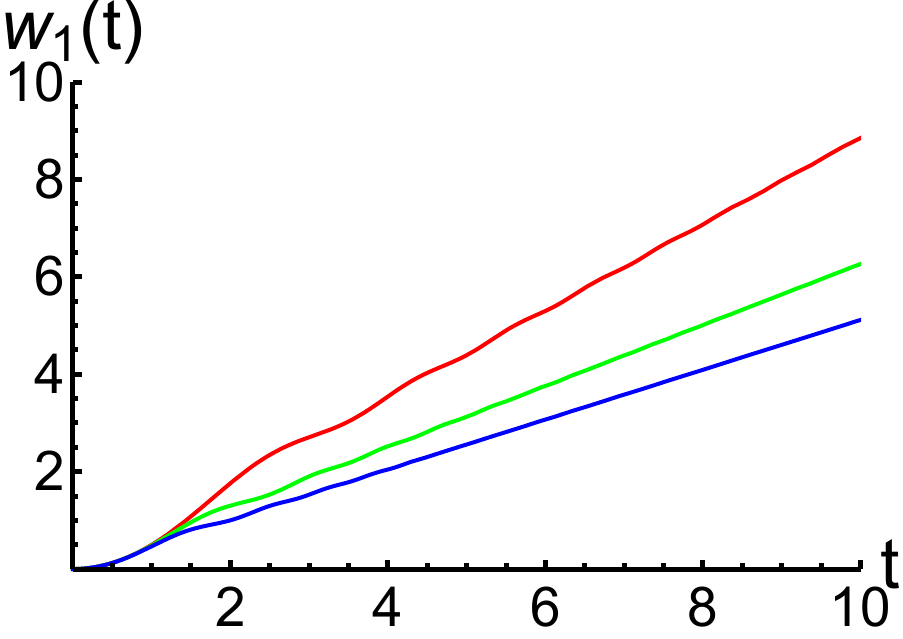}}
{\includegraphics[width=0.49\linewidth]{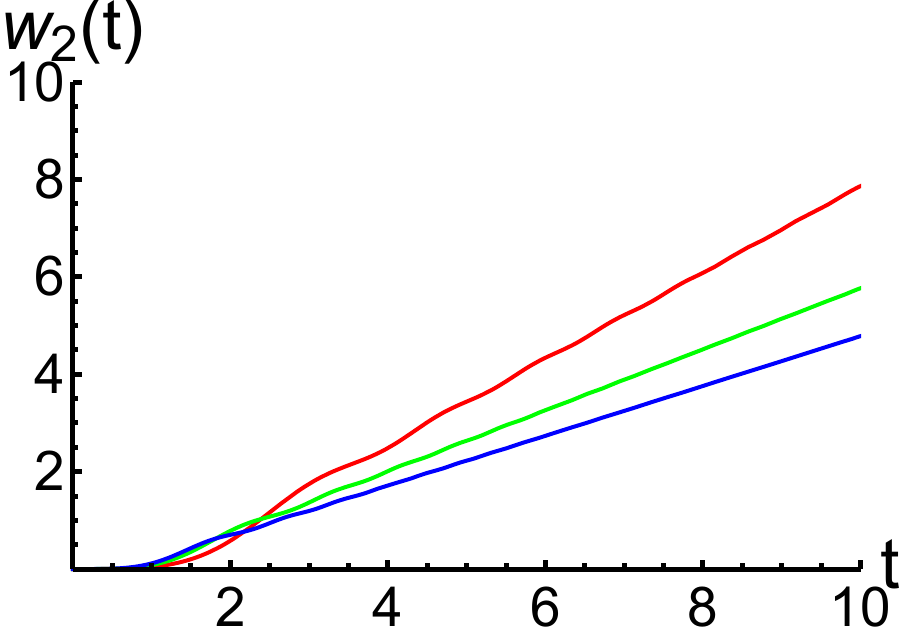}}
{\includegraphics[width=0.49\linewidth]{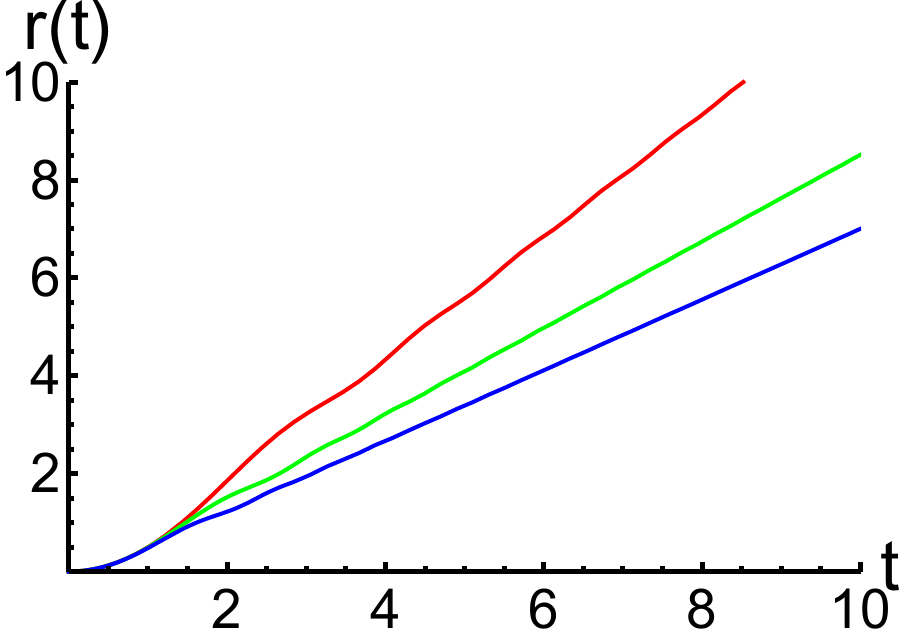}}
{\includegraphics[width=0.49\linewidth]{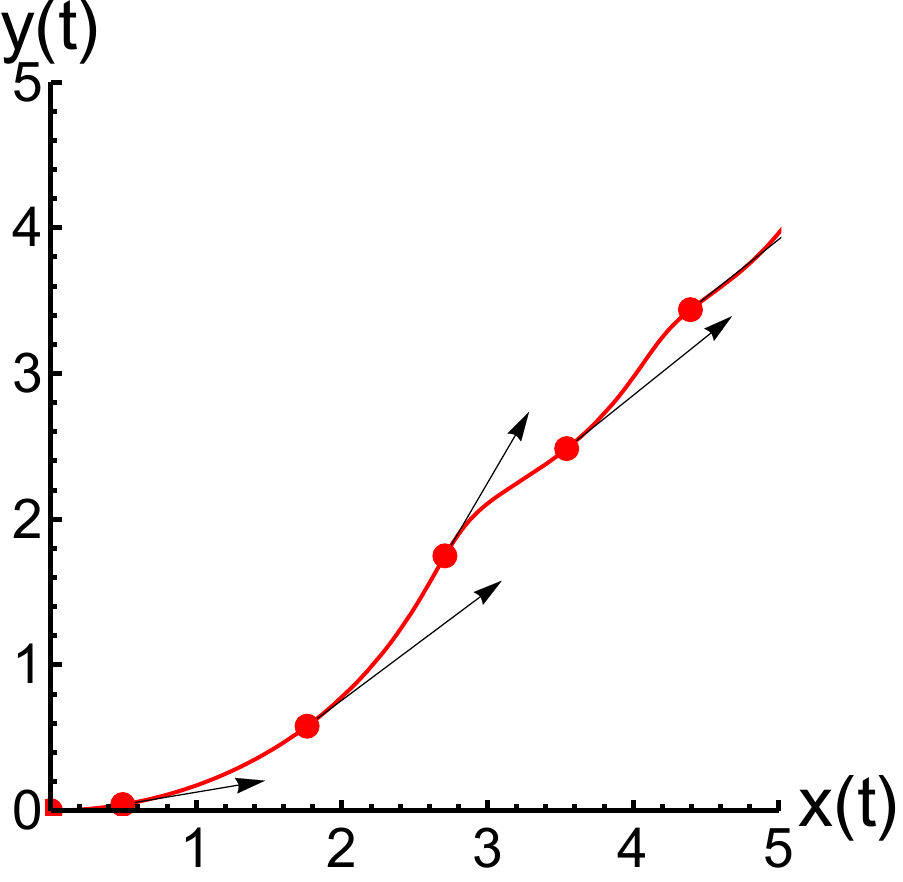}}
{\includegraphics[width=0.49\linewidth]{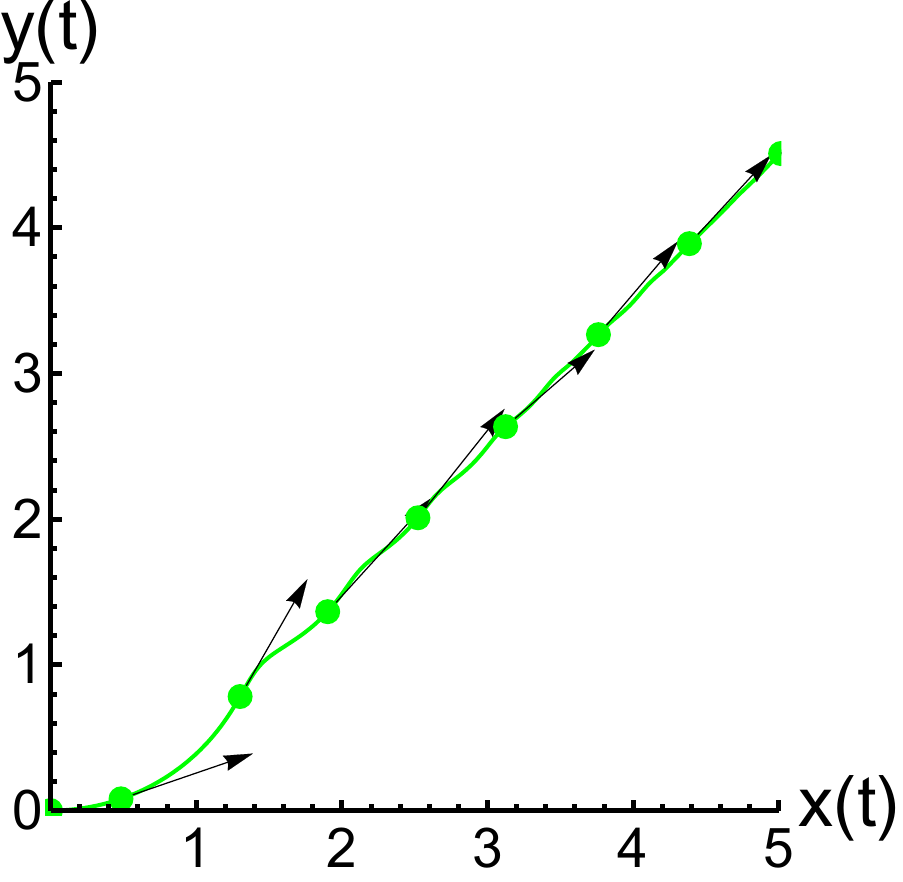}}
{\includegraphics[width=0.49\linewidth]{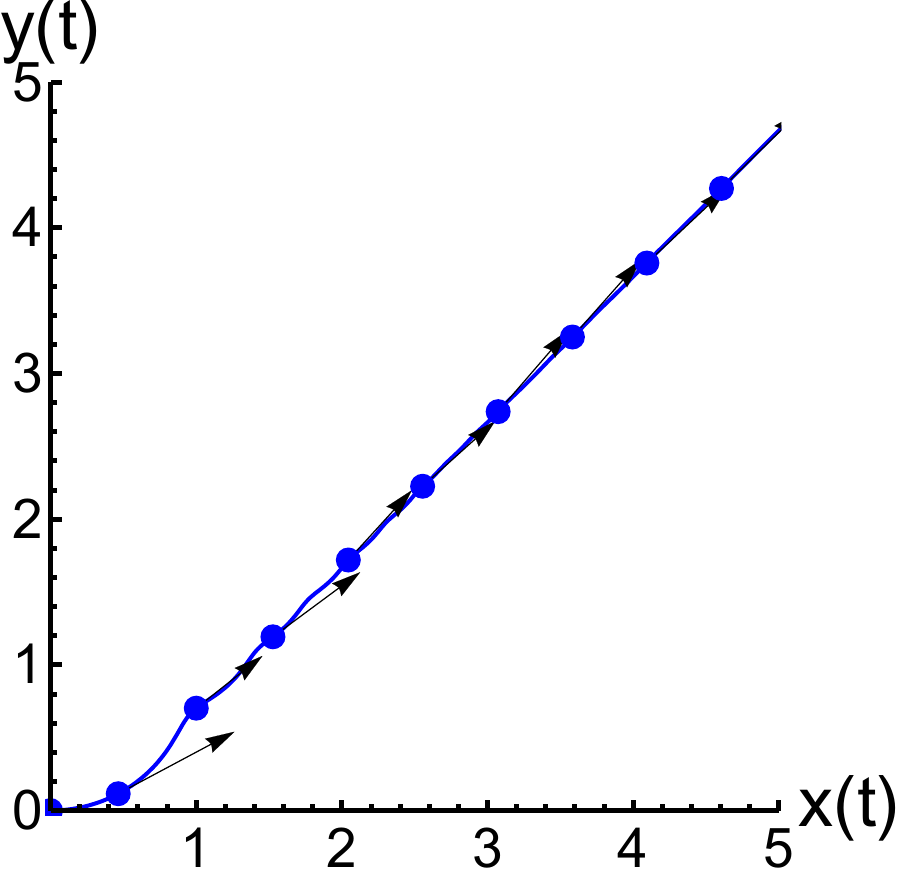}}
\caption{ Cartesian displacements $w_1\equiv x(t)$ (top left) and $w_2\equiv y(t)$ for a rigid particle
constrained to a plane under the action of a constant force acting as described in the text
for $K=1,2,3$.
The position on the surface, $\mathrm{r}(t)=\sqrt{x^2+y^2}$, with respect to the origin at time $t$ (middle left).
The fluctuations in direction of the speed (black arrows) in the course of motion for the same values of $K$ (middle right and bottom).
%The position on the surface, $\mathrm{r}(t)=\sqrt{x^2+y^2}$, with respect to the origin at time $t$ (bottom right).
Both force and mass are taken as unity.}
\label{fig7-8}
\end{figure}
and using $\ddot{x}$ from the second equation in the derivative of the first one,
we are led to the nonhomogeneous third-order linear differential equation
\begin{equation}\label{3dorder}
\dddot{w}+K^2t^2 \dot{w}-K^2tw=Kt~.
\end{equation}
This equation has the following linear independent solutions,
\begin{equation}
w_1(t) = x(t)~, \,\, w_2(t) = y(t)~, \,\, w_3(t) = \sqrt{\frac{\pi}{K}}\,t-\frac{1}{K}~.\label{9c}
\end{equation}
The first two of them are jerked with intermingled jerks given in (\ref{eq6}) and are just the cartesian displacements given in (\ref{eq3}) and plotted in
Fig.~\ref{fig7-8}. The third linear independent solution is a non jerked, degenerate solution, since it is also a solution of the simpler first-order linear equation
\begin{equation}\label{eq12}
K^2t^2 \dot{w}-K^2tw=Kt~.
\end{equation}
This solution is discarded because of the initial conditions of the motion.

%\medskip

The importance of the third order differential equation resides in its usage
as a (decoupled) definition of the jerks which can be calculated from $\dddot{w}_1=K^2t(w_1-t \dot{w}_1)=-Kt\sin(Kt^2/2)$ and
$\dddot{w}_2=K^2t(w_2-t \dot{w}_2)=Kt\cos(Kt^2/2)$, respectively.

%\medskip

\section{Conclusion}

In the planar motion of a rigid body with cartesian velocity components expressed through the Fresnel integrals
the speed tends quickly to a constant value affected by small sinc ripples whose amplitudes are additionally damped by the $K$ parameter.
Consequently, there is no surprise that also the trajectory corresponds to a planar motion of almost uniform velocity with only some small undulations. However, these undulations are important as they reveal the jerked features of the motion which are determined by the third-order nonhomogeneous linear differential equation obtained in this paper.


\begin{thebibliography}{10}
\bibitem{ratw}
E.L.~Starostin, R.A.~Grant, G.~Dougill, G.H.M. van der Heijden, V.G.A. Goss,
The Euler spiral of rat whiskers, \textit{Sci. Adv.} \textbf{6} (2020) eaax5145, https://doi.org/10.1126/sciadv.aax5145

 \bibitem{op}
 L.~Bartholdi, A.G.~Henriques, Orange peels and Fresnel integrals, \textit{Math. Intelligencer} \textbf{34} (2012) 1-3,  	https://doi.org/10.1007/s00283-012-9304-1

\bibitem{Ferris}
	A.V.~Ferris-Prabhu, On the appearance of Fresnel's integrals in dynamics, \textit{Am. J. Phys.} \textbf{38} (1970) 1356-1357,
 https://doi.org/10.1119/1.1976105

\bibitem{Hecht}
  E.~Hecht, \textit{Optics}, Fifth Edition, (Pearson Education Limited, London, 2017) 457-541

    \bibitem{Fowles}
	G.R.~Fowles, \textit{Introduction to Modern Optics}, Second Edition, (Dover Publications Inc., New York, 1975) 106-147
	
		\bibitem{Rossi}
	B.~Rossi, \textit{Ottica}, (Masson Italia Editori, Milan, 1977) 178-260
	
	\bibitem{Goodman}
	J.W.~Goodman, \textit{Introduction to Fourier Optics}, Second Edition, (McGraw-Hill Companies, 1996) 63-90

\bibitem{Iizuka}
K.~Iizuka, \textit{Engineering Optics}, Third Edition, (Springer, New York, 2008) 53-100, https://doi.org/10.1007/978-0-387-75724-7

\bibitem{rme}
    H.C.~Rosu, S.C.~Mancas, E.~Flores-Gardu\~no, Riccati parametric deformations of the Cornu spiral,
    \textit{Z. Naturforsch. A} \textbf{73} (2018) 479-484,  https://doi.org/10.1515/zna-2018-0111
	
    \bibitem{B1910}
    G.D.~Birkhoff, On the solutions of ordinary linear homogeneous differential equations of the third order,
    \textit{Ann. Math.} {\bf 12} (1911) 103-127,  https://doi.org/10.2307/2007241

\end{thebibliography}
\end{document}